\documentclass{article}[12pt]
\usepackage {epsfig,subfigure}

\newcommand{\pip}{\ensuremath{\pi^+\,}}
\newcommand{\pim}{\ensuremath{\pi^-\,}}

\newcommand{\etap}{\ensuremath{\eta'\,}}

\newcommand{\cm}{\ensuremath{\,{\rm cm}}}

\newcommand{\s}{\ensuremath{\,{\rm s}}}

\newcommand{\pb}{\ensuremath{\,{\rm pb}}}

\def\Fr{Frascati}
\def\fr{\rlap{\kern.2ex\up a}}
\def\Ka{Karlsruhe}
\def\ka{\rlap{\kern.2ex\up b}}
\def\Le{Lecce}
\def\le{\rlap{\kern.2ex\up c}}
\def\N{Napoli}
\def\n{\rlap{\kern.2ex\up d}}
\def\BE{Beer-Sheva}
\def\be{\rlap{\kern.2ex\up e}}
\def\Pi{Pisa}
\def\pI{\rlap{\kern.2ex\up f}}
\def\Ra{Roma I}
\def\ra{\rlap{\kern.2ex\up g}}
\def\Rb{Roma II}
\def\rb{\rlap{\kern.2ex\up h}$\,$}
\def\Rc{Roma III}
\def\rc{\rlap{\kern.2ex\up i}}
\def\su{\rlap{\kern.2ex\up j}}
\def\T{Trieste/Udine}
\def\t{\rlap{\kern.2ex\up k}}
\def\hsa{ \ }
\def\aff#1{Dipartimento di Fisica dell'Universit\`a e Sezione INFN, #1, Italy.}
\def\hsb{\hskip 2.8mm}

\title{
\begin{flushright}
\small{
Contributed paper to Lepton Photon 2001 \\
Rome, July 23-28.}
\end{flushright}
Detection of $\phi\to f_0$(980)$\gamma$, $\phi\to a_0$(980)$\gamma$
  into 5 
  photons with KLOE at DA$\Phi$NE} 
\date{ }
\author{The KLOE Collaboration}

\begin{document}
\maketitle
\def\ifm#1{\relax\ifmmode#1\else$#1$\fi}
\def\eps{\ifm{\epsilon}} \def\epm{\ifm{e^+e^-}}
\def\rep{\ifm{\Re(\eps'/\eps)}}  \def\imp{\ifm{\Im(\eps'/\eps)}}  
\def\DAF{DA$\Phi$NE}  \def\sig{\ifm{\sigma}}
\def\gam{\ifm{\gamma}} \def\to{\ifm{\rightarrow}}
\def\pip{\ifm{\pi^+}} \def\pim{\ifm{\pi^-}}
\def\po{\ifm{\pi^0}} 
\def\pic{\ifm{\pi^+\pi^-}} \def\pio{\ifm{\pi^0\pi^0}} 
\def\ks{\ifm{K_S}} \def\kl{\ifm{K_L}} \def\kls{\ifm{K_{L,\,S}}} 
\def\ksl{\ifm{K_S,\ K_L}} \def\ko{\ifm{K^0}}
\def\K{\ifm{K}} \def\LK{\ifm{L_K}}
\def\Kb{\ifm{\rlap{\kern.3em\raise1.9ex\hbox to.6em{\hrulefill}} K}}
\def\ab{\ifm{\sim}}  \def\x{\ifm{\times}}
\def\ff{$\phi$--factory}
\def\sta#1{\ifm{|\,#1\,\rangle}} 
\def\amp#1,#2,{\ifm{\langle#1|#2\rangle}}
\def\kob{\ifm{\Kb\vphantom{K}^0}}
\def\f{\ifm{\phi}}   \def\pb{{\bf p}}
\def\L{\ifm{{\cal L}}}  \def\R{\ifm{{\cal R}}}
\def\up#1{$^{#1}$}  \def\dn#1{$_{#1}$}
\def\etal{{\it et al.}}
\def\BR{{\rm BR}}
\def\radl{\ifm{X_0}}
\def\deg{\ifm{^\circ}} 
\def\th{\ifm{\theta}}
\def\To{\ifm{\Rightarrow}}
\def\ot{\ifm{\leftarrow}}
\def\fo{\ifm{f_0}} \def\epe{\ifm{\eps'/\eps}}
\def\pbrn{ {\rm pb}}  \def\cm{ {\rm cm}}
\def\mub{\ifm{\mu{\rm b}}} \def\s{ {\rm s}}
\def\RR{\ifm{{\cal R}^\pm/{\cal R}^0}}
\def\dt{ \ifm{{\rm d}t} } \def\dy{ {\rm d}y } \def\pbrn{ {\rm pb}}
\def\kp{\ifm{K^+}} \def\km{\ifm{K^-}}
\def\kkb{\ifm{\ko\kob}} 
\def\epe{\ifm{\eps'/\eps}}
\def\ppc{\ifm{\pi^+\pi^-}}
\def\ppo{\ifm{\pi^0\pi^0}}
\def\pppco{\ifm{\pi^+\pi^-\pi^0}}
\def\pppo{\ifm{\pi^0\pi^0\pi^0}}
\def\vare{\ifm{\varepsilon}}
\def\etap{\ifm{\eta'}}

\def\pt#1,#2,{#1\x10\up{#2}}

\def\B{Bari}
\def\b{\rlap{\kern.2ex\up a}}
\def\O{IHEP}
\def\o{\rlap{\kern.2ex\up b}}
\def\Fr{Frascati}
\def\fr{\rlap{\kern.2ex\up c}}
\def\Ka{Karlsruhe}
\def\ka{\rlap{\kern.2ex\up d}}
\def\Le{Lecce}
\def\le{\rlap{\kern.2ex\up e}}
\def\Mo{Moscow}
\def\mo{\rlap{\kern.2ex\up f}}
\def\N{Napoli}
\def\n{\rlap{\kern.2ex\up g}}
\def\BE{Beer-Sheva}
\def\be{\rlap{\kern.2ex\up h}}
\def\co{\rlap{\kern.2ex\up i}}
\def\Pi{Pisa}
\def\pI{\rlap{\kern.2ex\up j}}
\def\Ra{Roma I}
\def\ra{\rlap{\kern.2ex\up k}}
\def\en{\rlap{\kern.2ex\up l}}
\def\Rb{Roma II}
\def\rb{\rlap{\kern.2ex\up m}$\,$}
\def\Rc{Roma III}
\def\rc{\rlap{\kern.2ex\up n}}
\def\su{\rlap{\kern.2ex\up o}}
\def\T{Trieste}
\def\t{\rlap{\kern.2ex\up q}}
\def\V{Virginia}
\def\v{\rlap{\kern.2ex\up r}}
\def\Z{Associate member}
\def\hsa{ \ }

\baselineskip 13pt
A.~Aloisio\n,\hsa
F.~Ambrosino\n,\hsa
A.~Antonelli\fr,\hsa 
M.~Antonelli\fr,\hsa 
C.~Bacci\rc,\hsa
G.~Barbiellini\t,\hsa 
F.~Bellini\rc,\hsa
G.~Bencivenni\fr,\hsa 
S.~Bertolucci\fr,\hsa 
C.~Bini\ra,\hsa
% ...

\vglue 2mm

\def\aff#1{Dipartimento di Fisica dell'Universit\`a e Sezione INFN, #1, Italy.}

{\baselineskip=12pt
\parskip=0pt
\parindent=0pt
\def\hsb{\hskip 2.8mm}

\leftline{\b\hsb \aff{\B}}
\leftline{\o\hsb Permanent address: Institute of High Energy Physics of Academica Sinica, 
Beijing, China.}
\leftline{\fr\hsb  Laboratori Nazionali di Frascati dell'INFN, Frascati, Italy.}
\leftline{\ka\hsb  Institut f\"ur Experimentelle Kernphysik, Universit\"at \Ka,
Germany.}
\leftline{\le\hsb \aff{\Le}}
\leftline{\mo\hsb Permanent address: Institute for Theoretical and Experimental Physics, Moscow,
Russia.}
\leftline{\n\hsb Dipartimento di Scienze Fisiche dell'Universit\`a e 
Sezione INFN, \N, Italy.}
\leftline{\co\hsb Physics Department, Columbia University, New York, USA.}
\leftline{\pI\hsb \aff{\Pi}}
\leftline{\ra\hsb Dipartimento di Fisica dell'Universit\`a ``La Sapienza'' e Sezione INFN,
Roma, Italy}
\leftline{\en\hsb Dipartimento di Energetica dell'Universit\`a ``La Sapienza'', Roma, Italy.}
\leftline{\rb\hsb Dipartimento di Fisica dell'Universit\`a ``Tor Vergata'' e Sezione INFN,
Roma, Italy}
\leftline{\rc\hsb Dipartimento di Fisica dell'Universit\`a ``Roma Tre'' e Sezione INFN,
Roma, Italy}
\leftline{\su\hsb Physics Department, State University of New York 
at Stony Brook, USA.}
\leftline{\t\hsb \aff{\T}}
\leftline{\v\hsb Physics Department, University of Virginia, USA.}
 }

\begin{abstract}

The KLOE detector at DA$\Phi$NE, the Frascati $\phi$-factory, has collected
about 30 pb$^{-1}$ by the end of 2000, corresponding to about 90 millions
of $\phi$(1020) mesons produced. The five photon final state has been
exploited to study the rare decays of $\phi$ into $f_0$(980)$\gamma$ and
$a_0$(980)$\gamma$, with statistical accuracy never reached before.
The preliminary results for these branching ratios, from a data sample of
17 pb$^{-1}$, are the following: Br($\phi\to 
f_0$(980)$\gamma$)=(23.7$\pm$0.6$_{stat}$)$\times$ 10$^{-5}$, Br($\phi\to 
a_0$(980)$\gamma$)=(5.8$\pm$0.5$_{stat}$)$\times$10$^{-5}$. 
The systematic error is expected to be less than 10\%, a precise
evaluation is going on.
For the ratio of the branching ratios of  $\phi\to
f_0\gamma$ to $\phi\to a_0\gamma$ we find  4.1$\pm$0.4$_{stat}$.  

\end{abstract} 

\section{Introduction} 

In the last years there has been a revival of interest in the lowest-lying 
scalar mesons with masses below 1 GeV which has motivated the experimental 
effort going on with the KLOE detector\cite{kloe} at DA$\Phi$NE\cite{dafne} 
the Frascati $\phi$-factory, where these resonances can be studied in 
detail in $\phi$ radiative decays.   

The 1 GeV region is an interesting and challenging domain: on one side it
is 
far below the perturbative QCD regime, on the other side strict Chiral 
Perturbation Theory is not expected to make reliable predictions at these  
energy values where resonance effects can show up. This region has thus  
become an important test ground both for analyses based both on QCD 
sum-rules\cite{defazio} and for effective chiral 
models\cite{oset,bramon,black}.  

The controversial nature of the $f_0$(980), $a_0$(980)\cite{close} and the
poor 
knowledge of their properties adds further interest to this 1 GeV energy 
region: along the years several proposal have been suggested concerning the 
constitution of these scalars as complex q\=qq\=q states\cite{jaffe}, 
K\=K molecules\cite{weinstein} or ordinary q\=q mesons\cite{tornqvist}. 
A precise measurement of Br($\phi\to f_0\gamma$), 
Br($\phi\to a_0\gamma$) and of the ratio of these branching ratios can 
help in better establishing the nature of these mesons\cite{ratio}. 

In this paper we describe the preliminary results on the decays 
$\phi\to f_0\gamma$, $a_0\gamma\to$5$\gamma$ from the analysis of 17 
pb$^{-1}$, corresponding to about 50 millions of $\phi$ decays, from the 
data sample collected by KLOE in the year 2000.  
A peculiar aspect of these decays is that both the branching ratio and the 
resonance mass shape depend strongly on the meson structure. 
Therefore, the analysis follows a scheme independent as much as possible on  
the model implemented in the Montecarlo.  

\section{Selection criteria for five photon events} 

The events are characterized by the presence of five prompt 
photons, \rm{i.e.} photons coming from the interaction point (I.P.) of 
DA$\Phi$NE.  
These photons are detected as energy deposits in the calorimeter 
that satisfy the condition $t-r/c=0$, where $t$ is the arrival time, $r$ is
the distance of the cluster from the I.P. and $c$ is the speed of light.  
We define a photon to be ``prompt'' if $|t-r/c|<5\sigma_t$\cite{calo}. 
An acceptance angular region corresponding to the polar angle interval 
21$^o-$159$^o$ is defined for the prompt photons, in order to exclude the 
blind region around the beam-pipe.   
The processes that mainly contribute to the five photon final state are the 
following: 
 
\begin{itemize} 
\item[-] $\phi\to f_0\gamma\to\pi^0\pi^0\gamma$ 
\item[-] $\phi\to\rho^0\pi^0\to\pi^0\pi^0\gamma$ 
\item[-] $\phi\to a_0\gamma\to\eta\pi^0\gamma$ ($\eta\to\gamma\gamma$) 
\item[-] $\phi\to\rho^0\pi^0\to\eta\pi^0\gamma$ ($\eta\to\gamma\gamma$) 
\end{itemize} 
 
and the non resonant ones: 
 
\begin{itemize} 
\item[-] e$^+$e$^-\to\omega\pi^0\to\pi^0\pi^0\gamma$ 
\item[-] e$^+$e$^-\to\omega\pi^0\to\eta\pi^0\gamma$ ($\eta\to\gamma\gamma$) 
\end{itemize} 
 
Also the three and seven photon final states 
 
\begin{itemize} 
\item[-] $\phi\to\eta\gamma\to\gamma\gamma\gamma$ 
\item[-] $\phi\to\pi^0\gamma\to\gamma\gamma\gamma$  
\item[-] e$^+$e$^-\to\gamma\gamma(\gamma)$  
\item[-] $\phi\to\eta\gamma\to\pi^0\pi^0\pi^0\gamma$ 
\end{itemize} 
 
have to be taken into account, because they can simulate five photon events 
due to machine background and cluster splitting or merging of close photons
and loss of soft photons. 
 
The analyses described in this paper make use of a constrained fit 
ensuring kinematic closure of the events. 
The free parameters of the fit are: the three coordinates (x, y, z) of the 
impact point on the calorimeter, the energy and the time of flight for 
each photon coming from the I.P., the two energies of the beams and the 
three coordinates of the position of the I.P.. 
The analysis procedure adopted is the following:  
\begin{enumerate} 
        \item fully neutral events with exactly five prompt photons 
          are selected;  
        \item the kinematic fit is applied on these events a first 
              time with the constraints of the total energy and momentum 
              conservation and satisfying $t-r/c=0$  
              for each prompt photon; 
        \item the photon pairing is performed looking for the best 
          combination, by minimizing the appropriate $\chi^2$ for each of 
          the following hypotheses: $\pi^0\pi^0\gamma$, $\eta\pi^0\gamma$, 
          $\omega\pi^0$, $\eta\gamma\to\gamma\gamma\gamma$ and 
          $\pi^0\gamma$.     
        \item the kinematic fit is applied again to each of the best 
          combination constraining also the invariant masses of the photon 
          pairs assigned to $\pi^0$ and $\eta$, without any assumption 
          on the $f_0$ and $a_0$ mass.  
\end{enumerate}

\section{$\phi\rightarrow\pi^0\pi^0\gamma$} 
 
The main background channels, with the expected signal to background  
(S/B) ratio obtained from \cite{SND_f0n,SND_a0n,SND_wpn,PDG2000}, are
listed  
in Tab.~\ref{Tab:Eff_fon}. 
After applying the first kinematic fit, the best photon combination 
producing the two $\gamma\gamma$ pairs with $\pi^0$ mass is searched for 
with the additional condition $M_{\pi\pi}>700$ MeV (the expected $f_0$ 
mass region). According to our simulation, this procedure correctly assigns  
photons in 95\% of the cases. 
The analysis cuts to select $\phi\to f_0\gamma\to\pi^0\pi^0\gamma$
events are: 
 
\begin{enumerate} 
\item $\chi^2_{\pi\pi\gamma}/{\rm ndf} < 5$; 
\item $5\,\sigma$ cut on the reconstructed $\pi^0$'s mass (the output of
  the  
      first kinematic fit, without mass constraints, is used);
\item $\cos\psi_{\omega\pi}>0.4$, where $\psi$ is the angle between the  
      primary photon and the pion's flight direction in the $\pi^0\pi^0$  
      rest frame. This cut is performed to reduce the $\phi\to
      a_0\gamma$  
      and $\phi\to\rho^0\pi^0$ backgrounds; 
\item $\chi^2_{\omega\pi}/{\rm ndf}<3$ and a 
      $3\,\sigma$ cut on the reconstructed $\omega$ mass to veto
      $e^+e^-\to\omega\pi^0$; 
\item $E_{\rm tot}>900$ MeV and $175^{\circ}<\Delta\phi<185^{\circ}$ for 
      the two most energetic photons to reject
      $e^+e^-\to\gamma\gamma(\gamma)$;  
\item $3\,\sigma$ cut on the radiative 
      photon energy ($E_{\gamma\ {\rm rad}}\sim 500$ MeV) to reject
      $\phi\to\pi^0\gamma$.   
\end{enumerate} 

In order to evaluate the analysis efficiency as a function of $M_{\pi\pi}$,
the 
generated $f_0$ mass has been divided into 20 MeV bins. The total variation
of the efficiency does 
not exceed $30\%$ of the mean value, which is reported in
tab.~\ref{Tab:Eff_fon} together with the efficiencies for the background
channels.  
 
\begin{table}[ht] 
\begin{center} 
\begin{tabular}{|l|c|c|}\hline 
Decay channel  &  Natural S/B  &  Analysis efficiency               \\
\hline 
$\phi\to f_0\gamma\to\pi^0\pi^0\gamma$      &  ---   &  39.7\%      \\ 
$e^+e^-\to\omega\pi^0\to\pi^0\pi^0\gamma$   &  0.6   &   1.2\%      \\ 
$\phi\to\rho^0\pi^0\to\pi^0\pi^0\gamma$     &  3.7   &   4.9\%      \\ 
$\phi\to a_0\gamma\to\eta\pi^0\gamma\to\gamma\gamma\pi^0\gamma$   
                                            &  3.5   &   1.9\%      \\ 
$\phi\to\eta\gamma\to\pi^0\pi^0\pi^0\gamma$ &  0.02  &   
                                                 $5\times10^{-4}$   \\
                                                 \hline 
\end{tabular} 
\end{center} 
\caption{S/B ratios and analysis efficiencies for the 
$\phi\to f_0\gamma\to\pi^0\pi^0\gamma$ decay and related background.} 
\label{Tab:Eff_fon} 
\end{table} 
 
\begin{figure}[p] 
  \begin{center} 
    \mbox{\epsfig{file=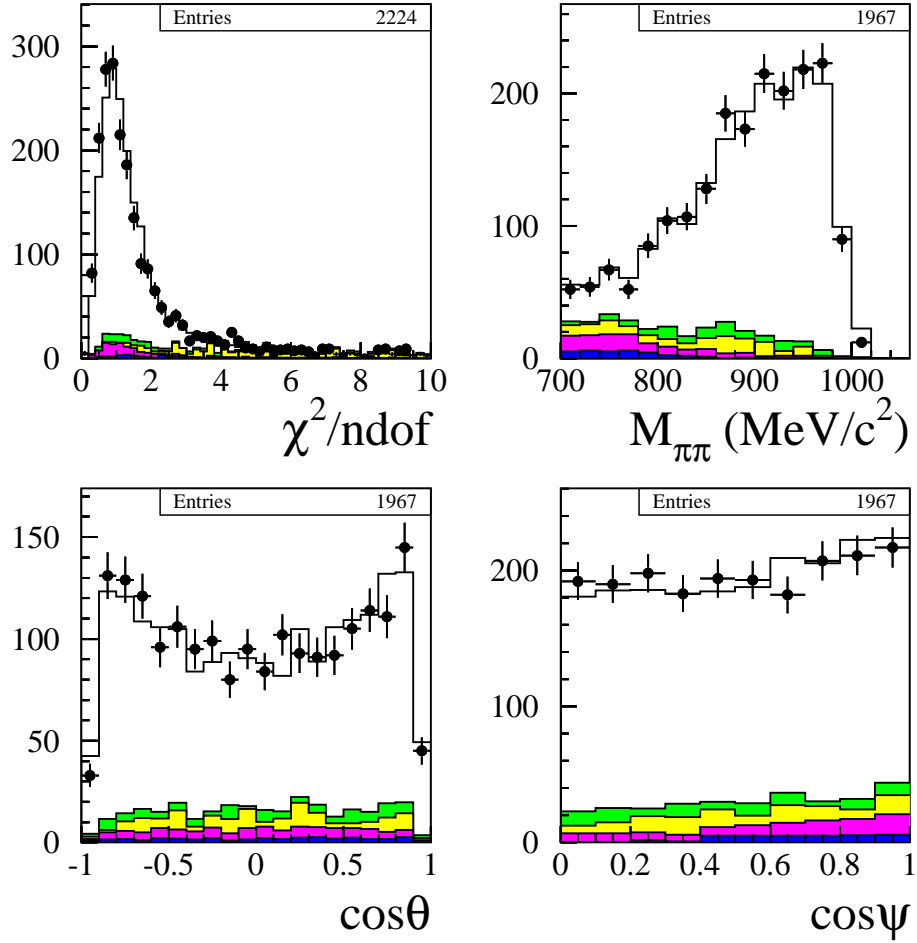,width=0.9\textwidth}} 
    \caption{$\phi\to f_0\gamma\to\pi^0\pi^0\gamma$ distributions:  
      $\chi^2/{\rm ndf}$ (up-left), $\pi^0\pi^0$ invariant mass  
      (up-right), polar angle (down-left) and $\cos\psi$ (down-right). 
      Black circles are data, solid histograms are the expected MC  
      signal+background spectra while the background contributions are 
      displayed in colors: green is $e^+e^-\to\omega\pi^0$, yellow is  
      $\phi\to\eta\gamma$, magenta is $\phi\to\rho\pi^0$ and blue is  
      $\phi\to a_0\gamma$. 
      The MC $f_0$ shapes have been obtained by weighting the $M_{\pi\pi}$  
      bin contents to reproduce the shape of the data.} 
    \label{Fig:f0n_distr} 
  \end{center} 
\end{figure} 
 
\begin{figure}[p] 
  \begin{center} 
    \mbox{\epsfig{file=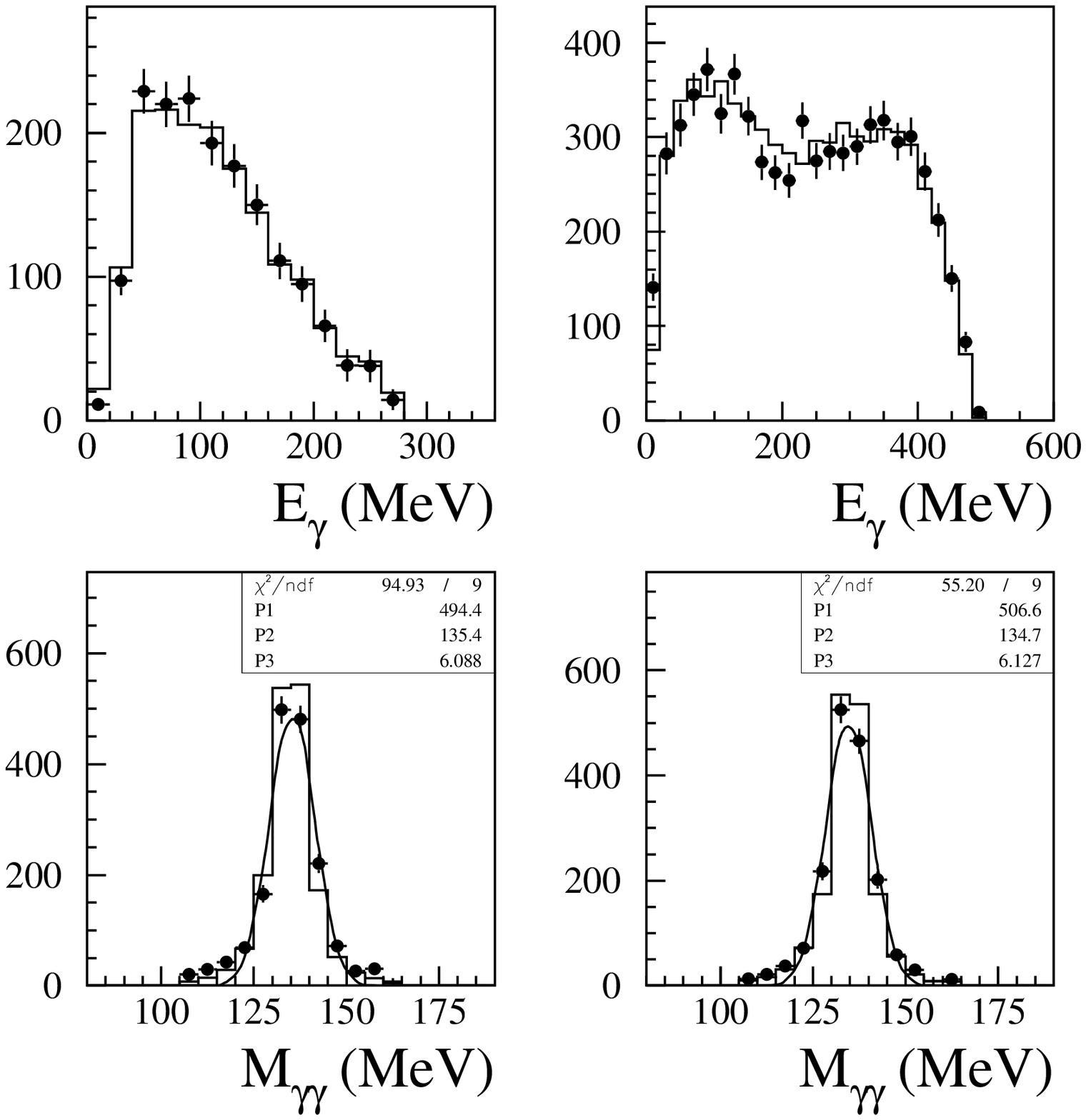,width=0.9\textwidth}} 
    \caption{$\phi\to f_0\gamma\to\pi^0\pi^0\gamma$ events: 
      energy distributions (up) [left: radiative photon, right: $\gamma$'s  
      from pions] and the two $\pi^0$ masses (down). 
      Black circles are data after background subtraction, solid 
      histograms are the expected spectra for the signal. 
      The MC $f_0$ shapes have been obtained by weighting the $M_{\pi\pi}$  
      bin contents to reproduce the shape of the data.} 
    \label{Fig:f0n_check} 
  \end{center} 
\end{figure} 
 
A fit procedure is in progress to evaluate the branching ratio and to
obtain the $f_0$  
mass shape by folding the available theoretical $dN/dM$ shapes with the  
reconstruction effects. 
Since this procedure is not completed and a reasonable $f_0$ mass shape was  
chosen in the Montecarlo, the branching ratio is evaluated using bin by bin
efficiency while data and Montecarlo 
distributions are compared as follows: 
 
\begin{enumerate} 
  \item $\phi\to f_0\gamma$ MC events are analyzed in slices of
    $M_{\pi\pi}$ 
        to obtain the various histograms shapes; 
  \item after subtracting the expected background from the observed
    $M_{\pi\pi}$ 
        distribution, for each $M_{\pi\pi}$  bin we evaluate the scale
        factor that, applied  
        on the MC $M_{\pi\pi}$ shape, reproduces data; 
  \item these scale factors are then applied to all MC ``sliced''
    histograms 
        which are then summed to produce the MC distributions to be
        compared 
        with the data. 
\end{enumerate} 
This implies that any effect of bad event reconstruction that gives rise to
a  
distorted $M_{\pi\pi}$ shape is neglected. We expect such contribution
mainly 
from bad photons' reconstruction ($\sim 4\div 5\%$) and wrong $\gamma$'s
pairing. 
The data--Montecarlo comparison for the most relevant distributions after
this  
procedure is shown in Fig.~\ref{Fig:f0n_distr}. The behaviour of the energy 
spectra and of the pion masses after the background subtraction is also 
reported (Fig.~\ref{Fig:f0n_check}).  
 
Applying the analysis to the 17 pb$^{-1}$, 1967 events survive while the  
Montecarlo expected background contribution is $305\pm 13_{\rm stat}\,$. 
The corresponding branching ratio has been evaluated  
by normalization with respect to the  
$\phi\to\eta\gamma\to\gamma\gamma\gamma$  events, using the
PDG\cite{PDG2000} value for the
Br($\phi\to\eta\gamma\to\gamma\gamma\gamma$).    
Neglecting the interference between the signal and   
$\phi\to\rho\pi^0\to\pi^0\pi^0\gamma$, the final branching ratio is: 
 
\begin{equation} 
Br(\phi\to f_0\gamma\to\pi^0\pi^0\gamma ) =  
    (\,7.9 \pm 0.2_{\rm stat}\,) \times 10^{-5}. 
\label{f0result} 
\end{equation} 
 
for $M_{\pi\pi}>700$ MeV. 
 
The systematic error is under study and it should not exceed $10\%$. 
  
\section{$\phi\rightarrow\eta\pi^0\gamma$} 
 
The $\phi\rightarrow\eta\pi^0\gamma$ events, with $\eta\to\gamma\gamma$,  
have been considered to study the $\phi\to a_0\gamma$ decay. 
The corresponding S/B ratio of the signal and the main background channels 
are reported in tab.\ref{rates}. 
 
\begin{table}[ht] 
  \begin{center} 
    \begin{tabular}{|c|c|c|} \hline 
Process                                    & Natural S/B  & Final
efficiency \\  
\hline 
$\phi\to a_0\gamma\to\eta\pi^0\gamma$       & --  &      27.2\% \\ 
$\phi\to\rho^0\pi^0\to\eta\pi^0\gamma$       & 5.3  &      27.1\% \\ 
e$^+$e$^-\to\omega\pi^0\to\eta\pi^0\gamma$      & 71   &      25.8\% \\ 
\hline 
e$^+$e$^-\to\omega\pi^0\to\pi^0\pi^0\gamma$     & 0.14 &
3.5$\times$10$^{-3}$ \\ 
$\phi\to\rho^0\pi^0\to\pi^0\pi^0\gamma$      & 1    &      5.0\%   \\ 
$\phi\to f_0\gamma\to\pi^0\pi^0\gamma$ & 0.27 &
1.4--3.0$\times$10$^{-3(*)}$ \\ 
\hline  
$\phi\to\eta\gamma\to\gamma\gamma\gamma$     & 6.1$\times$10$^{-3}$ & 
3.3$\times$10$^{-6}$ \\  
$\phi\to\eta\gamma\to\pi^0\pi^0\pi^0\gamma$  & 7.5$\times$10$^{-3}$ &
5.4$\times$10$^{-4}$ \\ 
    \hline 
    \end{tabular} 
    \caption{S/B ratios and efficiencies for $\eta\pi^0\gamma$ decay and 
      related background. $~~~~~~~~~~~~~~~~~~~~~~~~~~~~~~~~~~~~$
      $~~~~~~~~~~~~~^{(*)}$Depending on the $f_0$ shape.}    
    \label{rates} 
  \end{center} 
\end{table} 
 
\begin{figure}[p] 
\begin{tabular}{cc} 
(a)                    & (b)                        \\ 
\epsfig{file=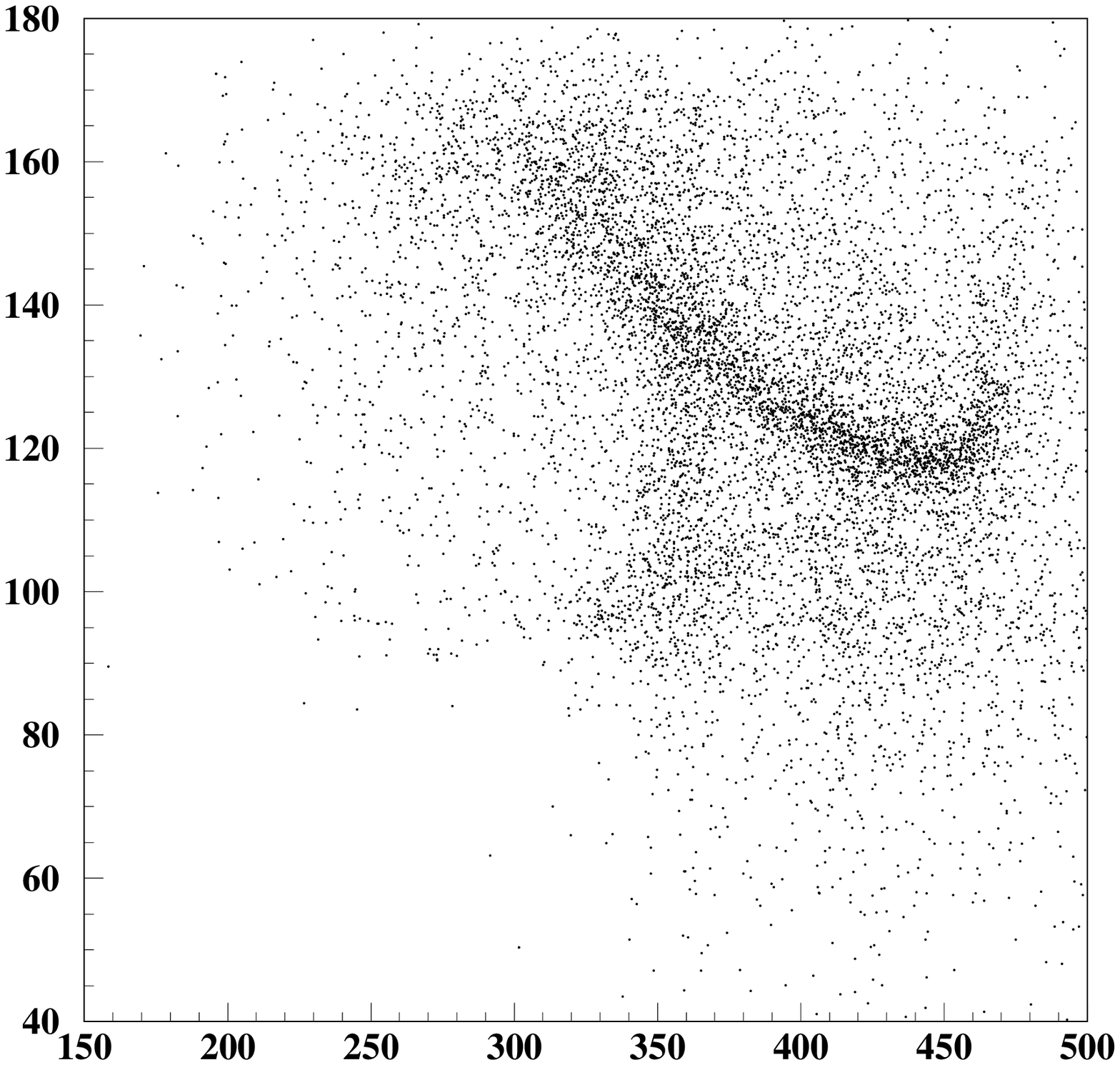, width=.48\textwidth} & \epsfig{file=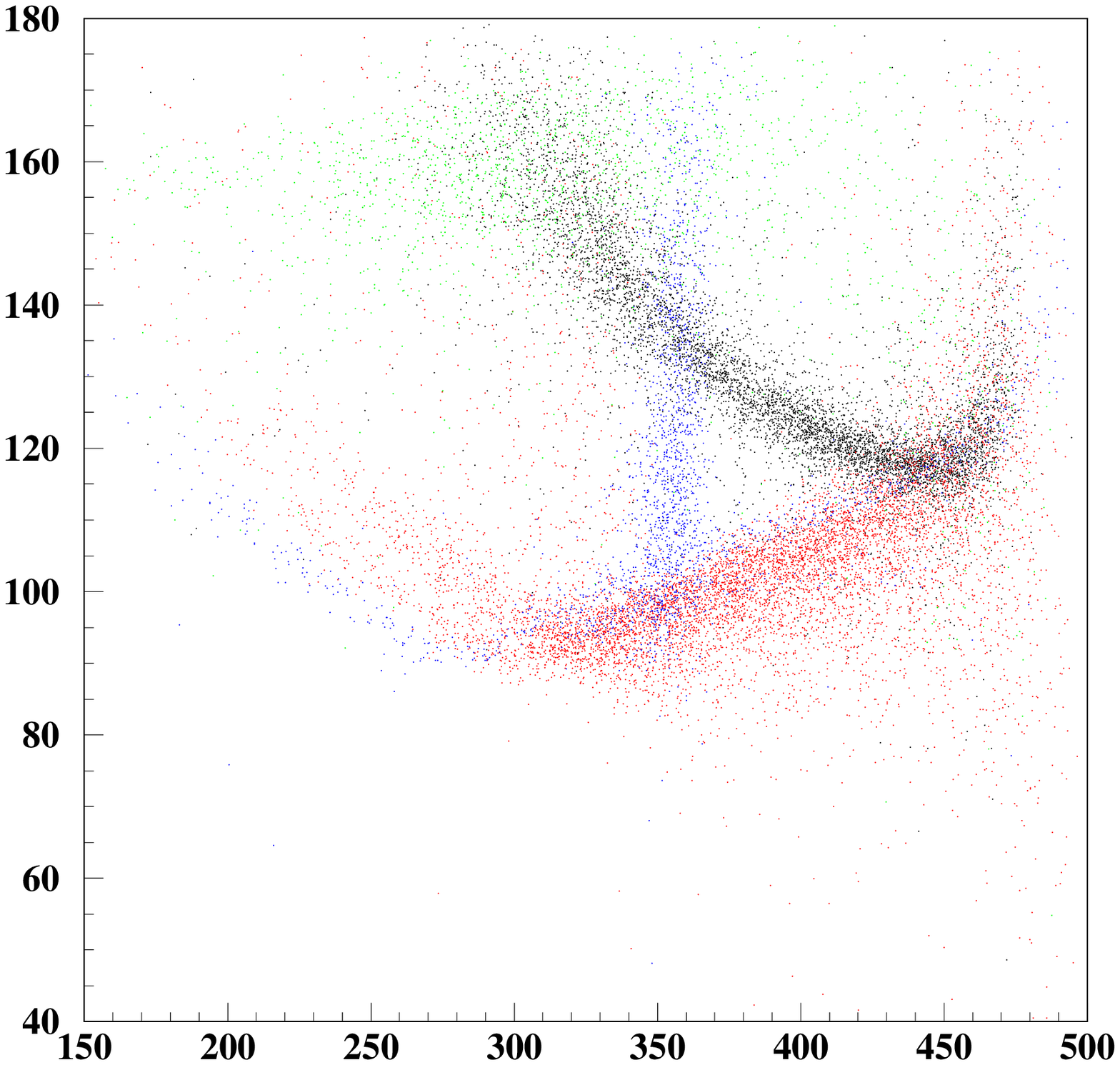,
  width=.48\textwidth} \\ 
$\psi$(deg) vs E$_{\gamma}$ (MeV) & $\psi$(deg) vs E$_{\gamma}$ (MeV) \\ 
(c)                    & (d)                        \\ 
\epsfig{file=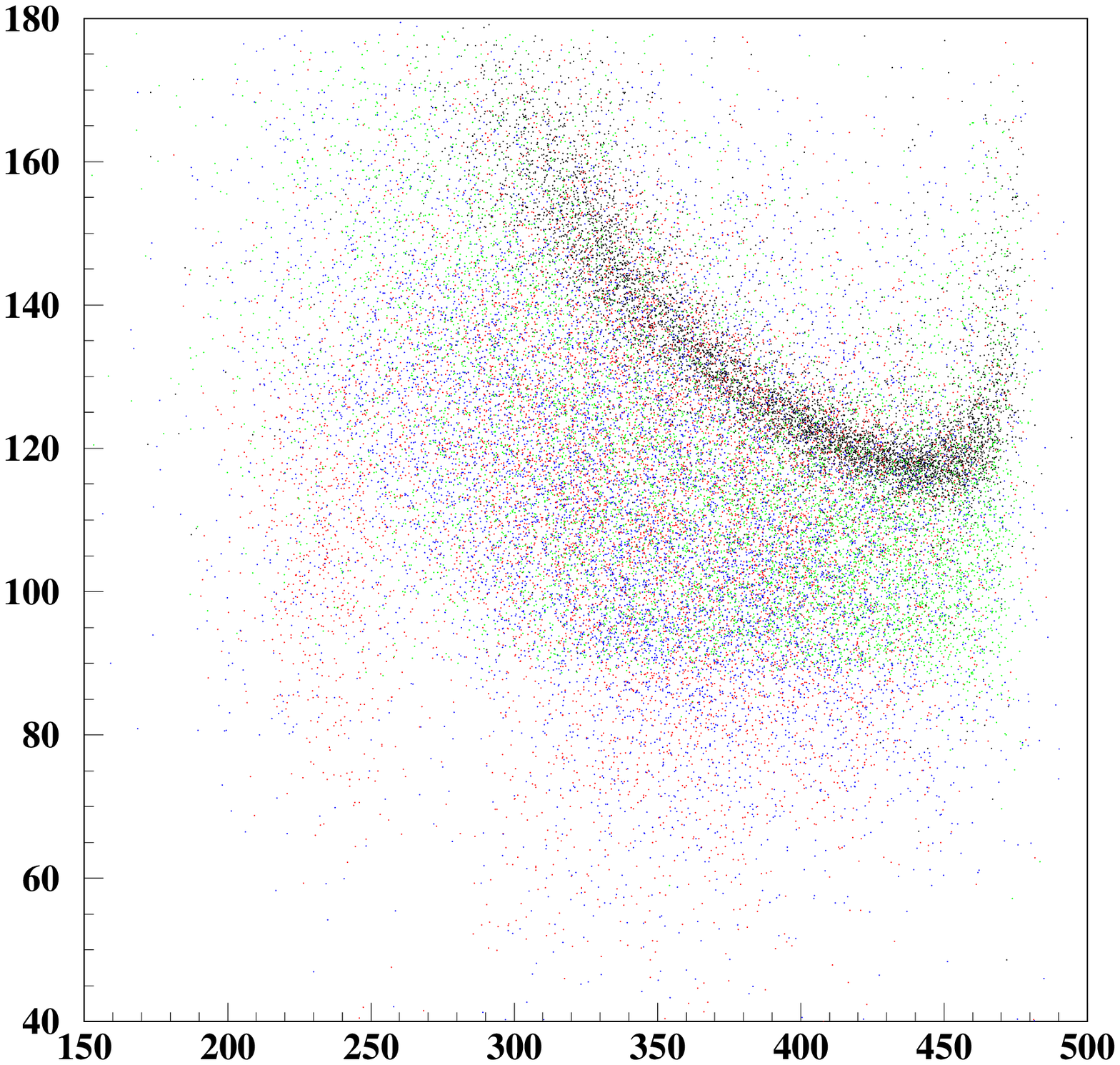, width=.48\textwidth} & \epsfig{file=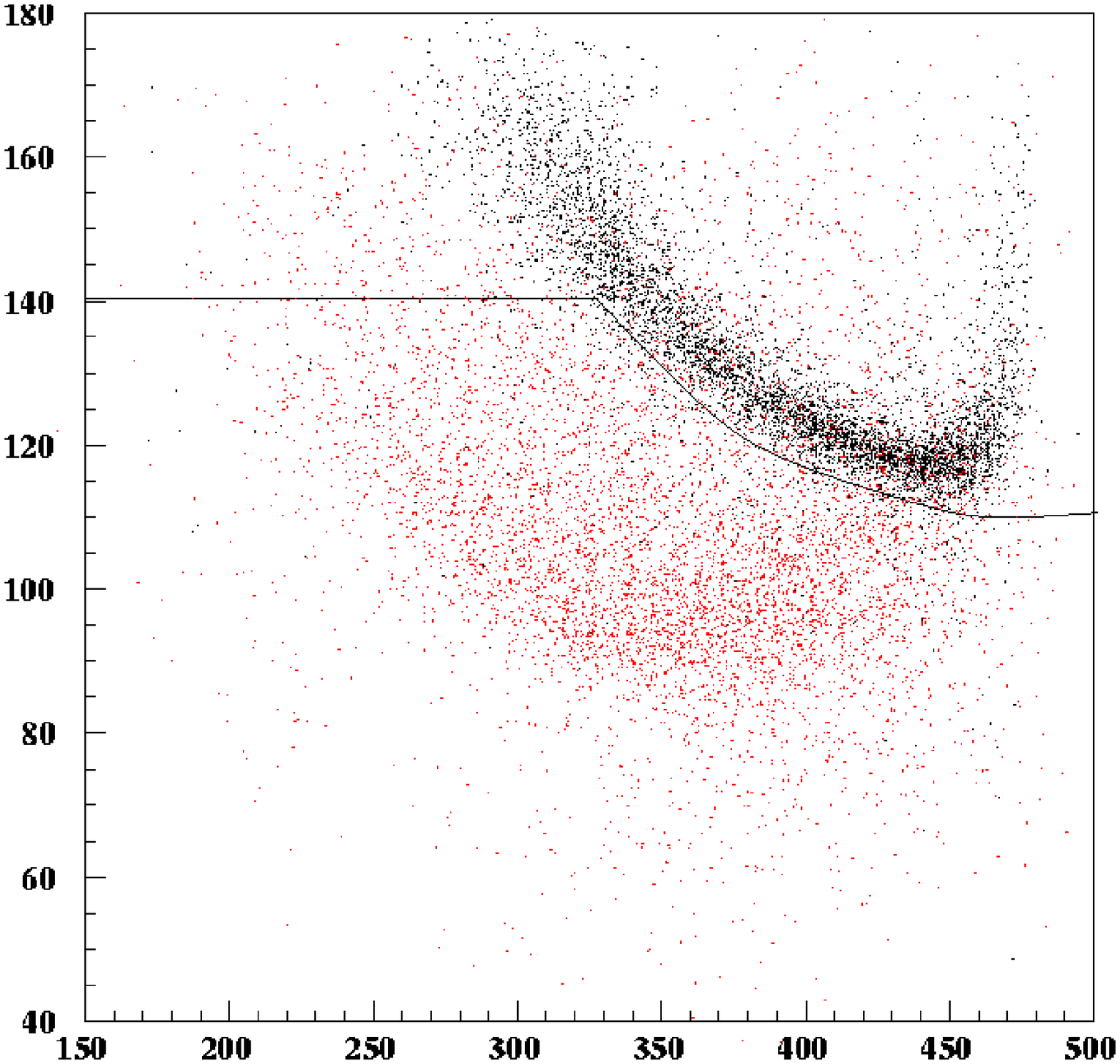,
  width=.48\textwidth} \\ 
$\psi$(deg) vs E$_{\gamma}$ (MeV) & $\psi$(deg) vs E$_{\gamma}$ (MeV) \\ 
\end{tabular} 
\caption{$\psi$ vs E$_{\gamma}$ - (a) Data; (b) MC: black - 
    $\omega\pi^0\to\pi^0\pi^0\gamma$, red - a$_0\gamma\to\eta\pi^0\gamma$, 
    green - f$_0\gamma\to\pi^0\pi^0\gamma$, blue - 
    $\eta\gamma\to\gamma\gamma\gamma$; (c) MC: black - 
    $\omega\pi^0\to\pi^0\pi^0\gamma$, red - 
    $\rho^0\pi^0\to\pi^0\pi^0\gamma$, green - 
    $\rho^0\pi^0\to\eta\pi^0\gamma$, blue - 
    $\omega\pi^0\to\eta\pi^0\gamma$; (d) MC: black - 
      $\omega\pi^0\to\pi^0\pi^0\gamma$; red - $\eta\pi^0\gamma$ flat (see 
    text); the accepted region is below the solid line.}    
\label{psi} 
\end{figure} 
 
\begin{figure}[p] 
  \begin{center} 
    \mbox{\epsfig{file=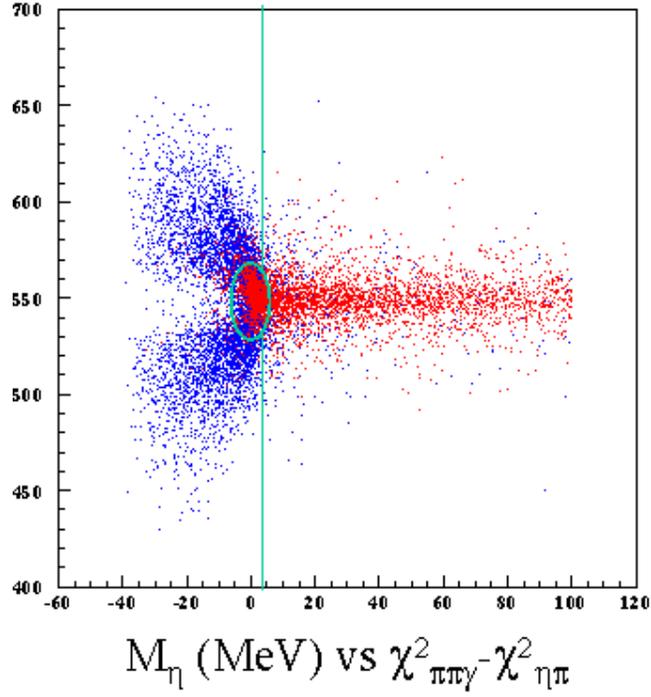,height=.4\textheight}} 
    \caption{$\eta$ reconstructed mass vs 
      $\chi_{\pi\pi\gamma}^2-\chi_{\eta\pi\gamma}^2$ for MC samples: red - 
      $\eta\pi^0\gamma$ flat; blue - $ \rho^0\pi^0\to\pi^0\pi^0\gamma$; the 
      applied cut is also shown.}  
    \label{elliptic} 
  \end{center} 
\end{figure} 
 
\begin{figure}[p] 
\begin{tabular}{cc} 
(a)                                       & (b) \\ 
\epsfig{file=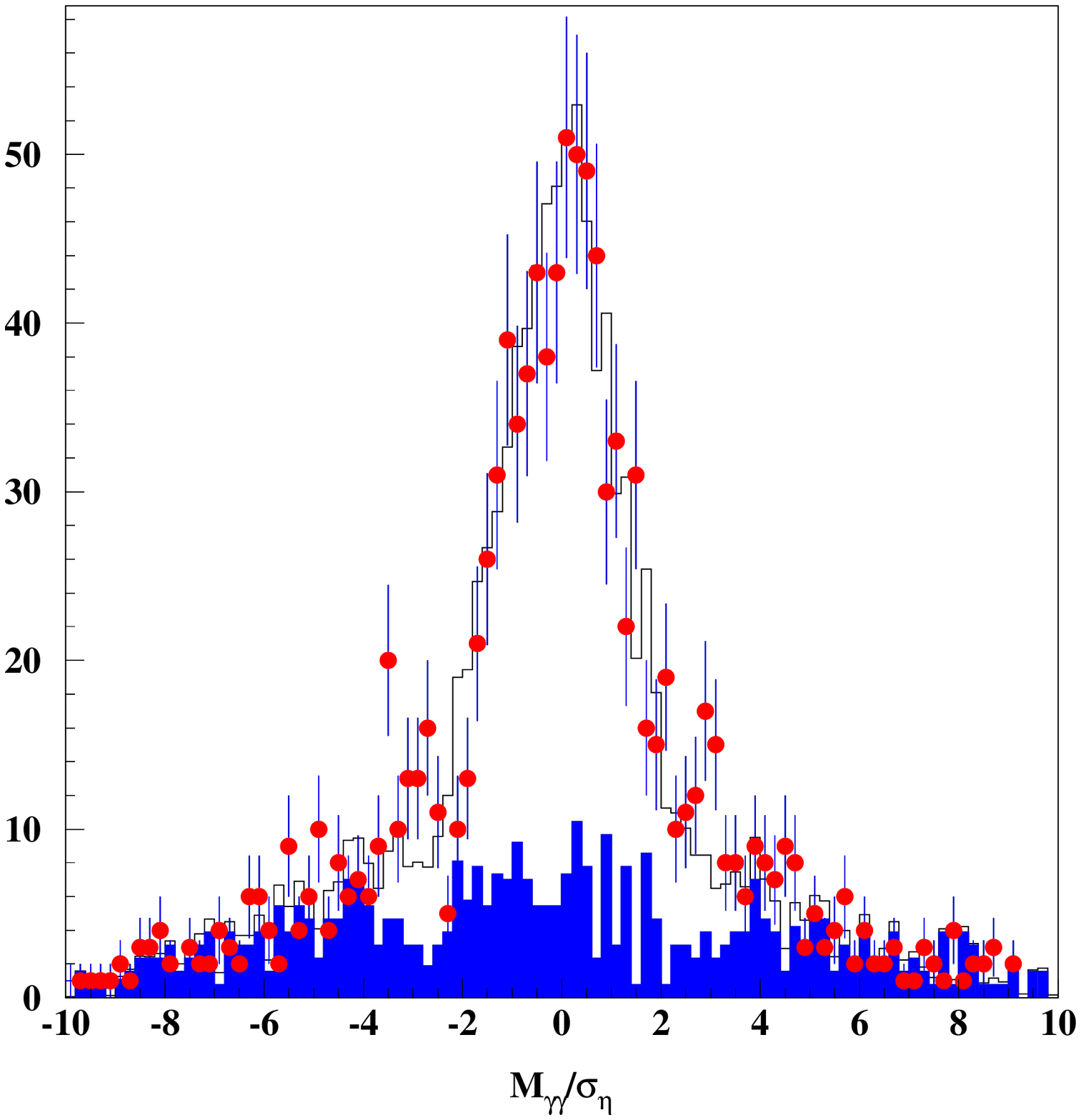, width=.48\textwidth} & \epsfig{file=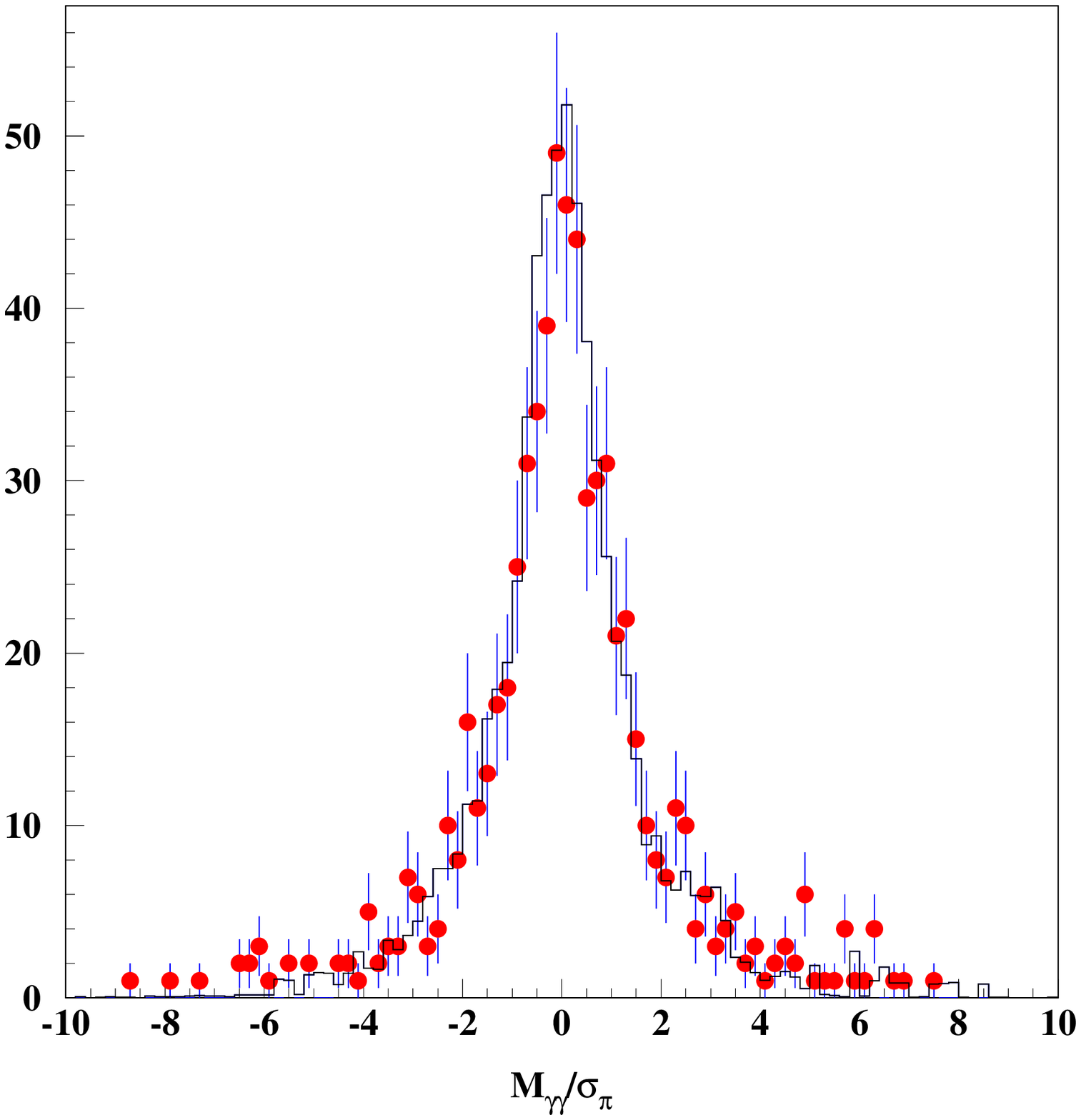, 
  width=.48\textwidth} \\ 
\end{tabular} 
\caption{(a) Reconstructed $\eta$ mass divided by the resolution; full 
  histogram: $\eta\gamma\to\pi^0\pi^0\pi^0\gamma$ (MC); (b) Reconstructed 
      $\pi^0$ mass divided by the resolution.}      
\label{erad} 
\end{figure} 
 
A first selection to reject e$^+$e$^-\to\gamma\gamma$($\gamma$) events has
been applied.  
Then the $\eta\gamma\to\gamma\gamma\gamma$ contamination is eliminated by
rejecting events with a two photon invariant  
mass near the $\eta$ mass together with another photon energy near 363 MeV.  
  
In order to reduce the $\pi^0\pi^0\gamma$ background, the photon 
combination obtained for the e$^+$e$^-\to\omega\pi^0\to\pi^0\pi^0\gamma$ 
hypothesis has been exploited. 
The correlation between the angle $\psi$ (angle between the non associated 
photon and the primary $\pi^0$ in the dipion rest frame), and the energy of 
the non associated photon is shown in fig.\ref{psi}.a,b for MC samples and 
in fig.\ref{psi}.c for data. 
From the comparison several contributions can be identified. 
In order to avoid introducing any bias in the unknown $a_0$ shape, the 
analysis cuts have been studied using a $\phi\to\eta\pi^0\gamma$ MC
sample 
with an almost flat $\eta\pi^0$ invariant mass. 
That sample is compared in fig.\ref{psi}.d with the dominant background;
also the applied cut is shown. 
 
\begin{figure}[p] 
\begin{tabular}{cc} 
(a)                                                 & (b) \\  
  \epsfig{file=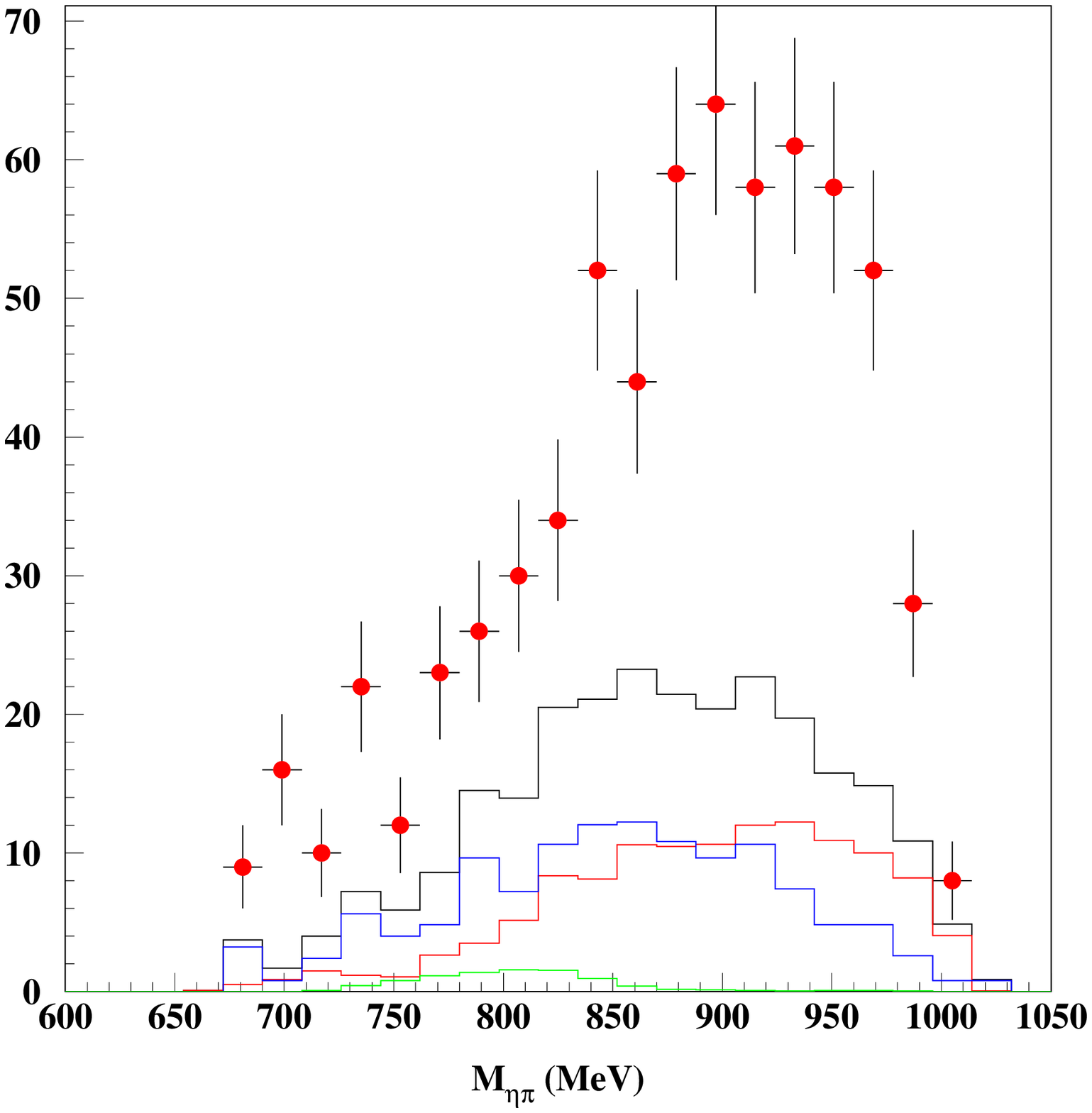, width=.48\textwidth} & 
  \epsfig{file=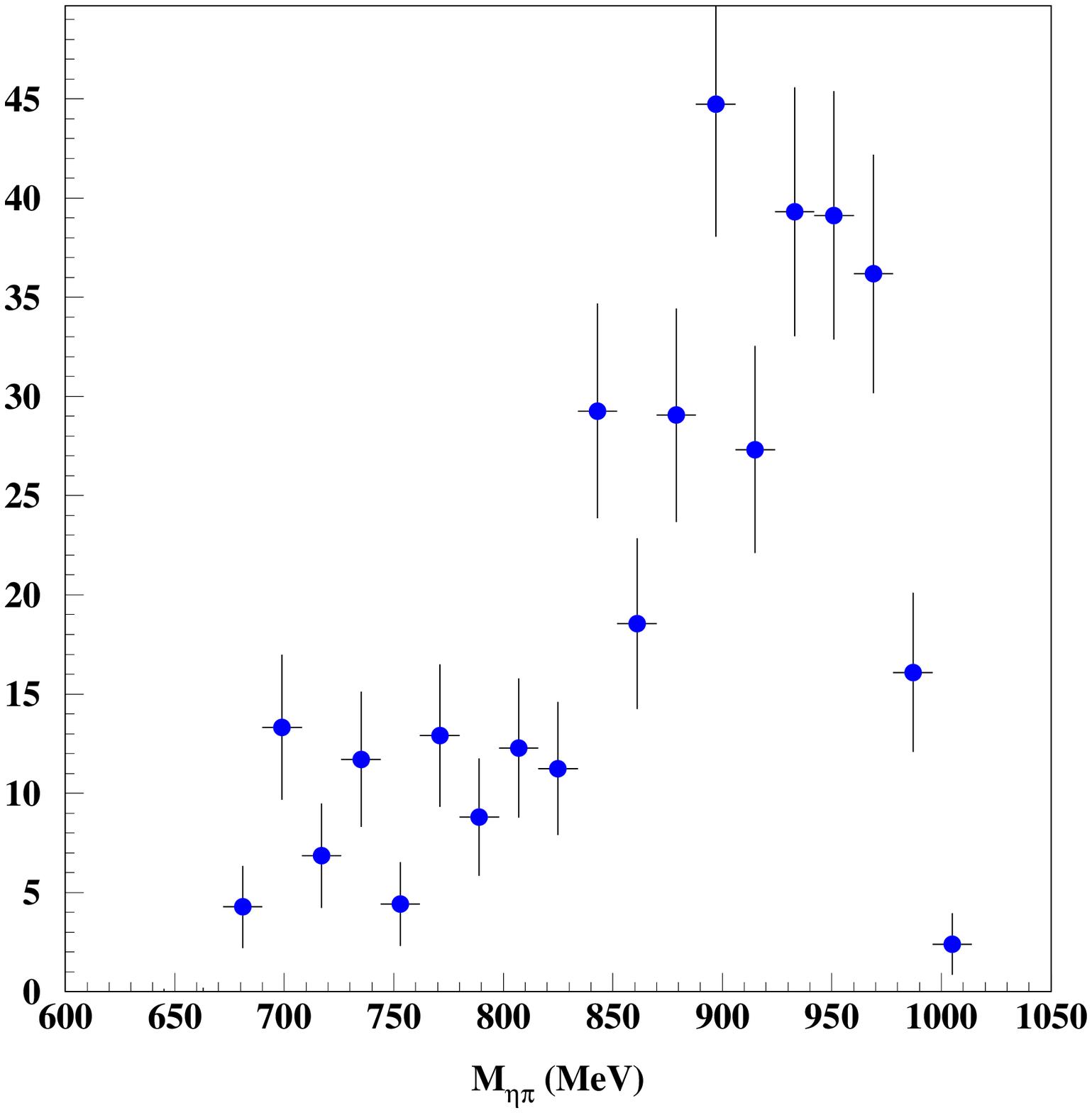, width=.48\textwidth} \\  
  \end{tabular} 
\begin{center}
\begin{tabular}{c}
  (c) \\
  \epsfig{file=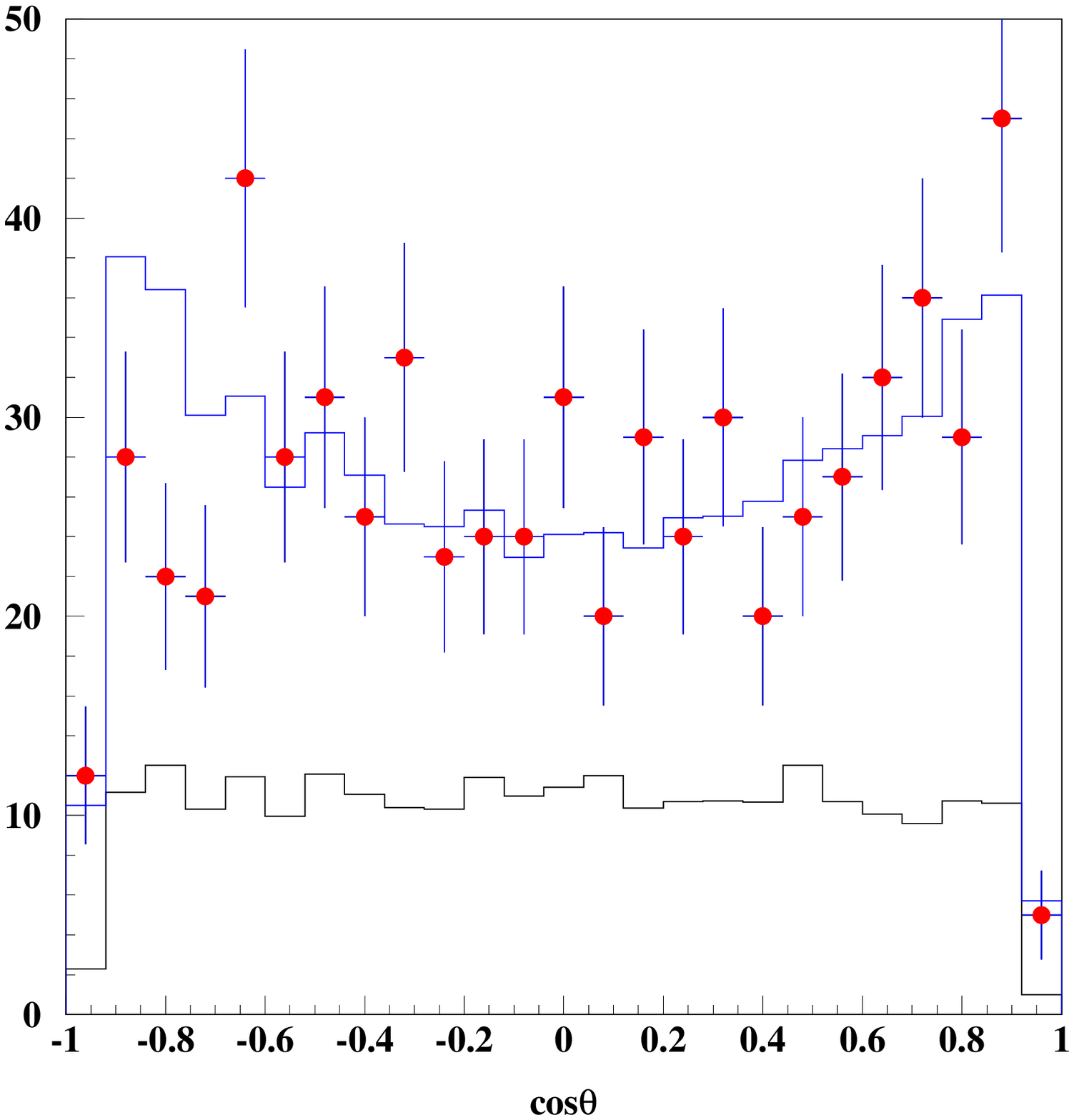, width=.48\textwidth} \\  
\end{tabular}
\end{center}
    \caption{(a) $\eta\pi^0$ invariant mass - points: data, histograms: 
      background spectra (MC), red - $\pi^0\pi^0\gamma$, 
      green - e$^+$e$^-\to\omega\pi^0\to\eta\pi^0\gamma$, blue - 
      $\eta\gamma\to\pi^0\pi^0\pi^0\gamma$, black - total; (b) $\eta\pi^0$
      invariant mass background subtracted; (c)  
      cos$\vartheta$ distribution - points: data, histograms: background, 
      signal plus background.} 
    \label{a0spectrum} 
\end{figure} 
 
Since the $\phi\to\rho^0\pi^0\to\pi^0\pi^0\gamma$ events populate the
same 
region of the scatter plot of fig.\ref{psi} as the signal, this 
contamination should be reduced by using different variables.  
The choosen ones are shown in the scatter plot of fig.\ref{elliptic}: the 
difference between the $\chi^2$ of the second fit in the hypotheses 
$\pi^0\pi^0\gamma$ and $\eta\pi^0\gamma$ and the reconstructed $\eta$
mass. 
The accepted region is defined by  
$\chi_{\pi^0\pi^0\gamma}^2-\chi_{\eta\pi^0\gamma}^2>$3 plus the internal 
part of the ellipse. 
 
Last contamination that has to be reduced is 
$\phi\to\eta\gamma\to\pi^0\pi^0\pi^0\gamma$.  
In fig.\ref{erad}.a is reported the invariant mass of the two photons 
associated to the $\eta$; the comparison with the MC shows that the 
background events mainly populate the tails of the distribution. 
A two standard deviation cut is then applied to reduce the $\eta\gamma$ 
contamination.  
The final efficiencies are reported in tab.\ref{rates}. 
 
At the end of the selection procedure 666 events survive.  
The expected number of background events is 253$\pm$11 mainly due to 
$\phi\to\rho^0\pi^0\to\pi^0\pi^0\gamma$ and  
$\phi\to\eta\gamma\to\pi^0\pi^0\pi^0\gamma$.   
The invariant mass of $\eta\pi^0$ is shown in fig.\ref{a0spectrum},
together with 
the cos$\vartheta$ distribution for the non associated photon. 
The corresponding branching ratio is: 
\begin{equation} 
Br(\phi\to\eta\pi^0\gamma)=(7.4\pm 0.5_{stat})\times 10^{-5} 
\label{a0result} 
\end{equation} 
The systematic error is still under study; according to preliminary 
evaluations it should not exceed 10\%. 
This result is in agreement, within the errors, with the values published
by the VEPP-2M experiments\cite{SND_a0n,a0cmd2} and with the KLOE result of  
the analysis of the 1999 statistics\cite{osaka}. 
 
The branching ratio (\ref{a0result}) includes the 
$\phi\to\rho^0\pi^0\to\eta\pi^0\gamma$ contribution.    
If the interference of this process with 
$\phi\to a_0\gamma\to\eta\pi^0\gamma$ is negligible\cite{achasov}, one
can 
subtract the $\rho^0\pi^0$ contribution. 
To evaluate the number of expected $\rho^0\pi^0$ events, the average of the 
very recent measurements of the Br($\rho^0\to\eta\gamma$) performed by the 
two VEPP-2M experiments\cite{rhovepp} has been used. 
This average value turns out to be 
Br($\rho^0\to\eta\gamma$)=(3.0$\pm$0.3)$\times$10$^{-4}$ (which is 
different from the PDG value\cite{PDG2000} 
Br($\rho^0\to\eta\gamma$)=(2.4$\pm$0.9)$\times$10$^{-4}$), from which 
follows an expected number of 86$\pm$9 $\rho^0\pi^0\to\eta\pi^0\gamma$
events. 
After the subtraction one obtains: 
\begin{equation} 
Br(\phi\to a_0\gamma\to\eta\pi^0\gamma)=(5.8\pm 0.5_{stat})\times
10^{-5}   
\label{a0subtracted} 
\end{equation}

\section{$\phi\to f_0\gamma$ to $\phi\to a_0\gamma$ ratio} 
 
From the results described in the previous sections a preliminary estimate 
of the ratio of the two branching ratios can be obtained. 
Assuming that $\pi^0\pi^0\gamma$ and $\eta\pi^0\gamma$ are the dominant 
decay modes of respectively $f_0$ and $a_0$ and using the values of 
eqs.(\ref{f0result}) and (\ref{a0subtracted}), one obtains (with 
Br($\phi\to f_0\gamma$)=3Br($\phi\to f_0\gamma\to\pi^0\pi^0\gamma$)): 
\begin{equation} 
\frac{Br(\phi\to f_0\gamma)}{Br(\phi\to a_0\gamma)}=4.1\pm 0.4_{stat} 
\end{equation} 
in reasonable agreement with the prediction given in \cite{ratio}. 

\section{Conclusions} 
 
Events with five prompt photon have been selected from a sample of 17 
pb$^{-1}$ of the data collected by KLOE in 2000, corresponding to  
about 50 millions of $\phi$ meson produced. 
From the analysis of the $\pi^0\pi^0\gamma$ channel, a 
Br($\phi\to
f_0$(980)$\gamma\to\pi^0\pi^0\gamma$)=(7.9$\pm$0.2$_{stat}$)$\times$  
10$^{-5}$ has been obtained. 
The result of the analysis of the $\eta\pi^0\gamma$ with 
$\eta\to\gamma\gamma$ channel, is:  
Br($\phi\to\eta\pi^0\gamma$)=(7.4$\pm$0.5$_{stat}$)$\times$10$^{-5}$. 
By subtracting the expected $\phi\to\rho^0\pi^0\to\eta\pi^0\gamma$ 
contribution, one can evaluate Br($\phi\to 
a_0$(980)$\gamma\to\eta\pi^0\gamma$)=(5.8$\pm$0.5$_{stat}$)$\times$10$^{-5}$. 
The ratio of the branching ratio $\phi\to f_0\gamma$ to
$\phi\to a_0\gamma$ = 4.1$\pm$0.4$_{stat}$ is obtained. \\ 
Work is in progress to fit the spectra and to extract the $f_0$ and $a_0$ 
resonance parameters.


\begin{thebibliography}{999} 
\bibitem{kloe} KLOE Collaboration, KLOE: a general purpose dcetector for 
  DA$\Phi$NE, LNF-92/019 (IR) (1992); KLOE Collaboration, The KLOE detector 
  - Technical Proposal, LNF-93/002 (IR) (1993). 
\bibitem{dafne} S. Guiducci et al., Proceedings of PAC99, New York, March
  1999. 
\bibitem{defazio} F. De Fazio and M. Pennington, hep-ph/0104289, June 
  2001.  
\bibitem{oset} E. Oset et al., Physics and detectors for DA$\Phi$NE - 
  Frascati, Nov.16-19,1999 Frascati Physics Series Vol.XVI (2000); . 
  J.A. Oller, hep-ph/0007349 (2000).  
\bibitem{bramon} A. Bramon et al., Phys. Lett. B 494 (2000), 221; 
  R. Escribano,  hep-ph/0012050; A. Bramon et al., hep-ph/0105179 (to be 
  published in Phys. Lett. B).  
\bibitem{black} D. Black et al., Phys. Rev. D61 (2000), 074001. 
\bibitem{close} F. Close, N. Isgur and S. Kumano, Nucl. Phys. B389 (1993) 
  513; See also N. Brown, F.E. Close in: L. Maiani, G. Pancheri, N. Paver
  (Eds.), 
    The Second Dafne Handbook, INFN-LNF, 1995, p.649. 
\bibitem{jaffe} R. L. Jaffe, Phys. Rev. D15 (1977) 267; M. Alford, 
  R.L. Jaffe, Nucl. Phys. B578 (2000) 367.  
\bibitem{weinstein} J. Weinstein and N. Isgur, Phys. Rev. Lett. 48 (1982)
  659 
\bibitem{tornqvist} N.A. T\"{o}rnqvist, Z. Phys. C68 (1995) 647 and
  references 
    therein.  
\bibitem{ratio} F.E. Close, A. Kirk, hep-ph/0106108 (2001). 
\bibitem{calo} M. Adinolfi et al.[The KLOE Collaboration], LNF-01/017 (P)
  (2001), submitted to Nucl. Intr. and Meth. 
\bibitem{SND_f0n} M.N.Achasov et al., Phys.~Lett.~B485 (2000), 349. 
\bibitem{SND_a0n} M.N.Achasov et al., Phys.~Lett.~B479 (2000), 53. 
\bibitem{SND_wpn} M.N.Achasov et al., Nucl.~Phys.~B569 (2000), 158. 
\bibitem{PDG2000} D.E.Groom   et al., The European Physical Journal C15 
  (2000), 1. 
\bibitem{a0cmd2} R.R. Akhmetshin et al., Phys. Lett. B462 (1999), 280.  
\bibitem{osaka} The KLOE Collaboration, hep-ex/0006036.  
\bibitem{achasov} N.N. Achasov and V.V. Gubin, Phys. Rev. D63 (2001) 
  094007. 
\bibitem{rhovepp} M.N. Achasov et al., JETP Lett. 72 (2000), 282; 
  R.R. Akhmetshin et al., Phys. Lett. B509 (2001), 217. 
 
\end{thebibliography}
\end{document}